\documentclass[sigconf,balance=false]{acmart}
\usepackage{popets}

\setcopyright{none} 
\renewcommand\footnotetextcopyrightpermission[1]{} 
\copyrightyear{YYYY}

\acmISBN{}
\acmConference{Proceedings on Privacy Enhancing Technologies}
\begin{document}

\title{Location Privacy in B5G/6G: Systematization of Knowledge}


\author{Hannah B. Pasandi}
\orcid{}
\affiliation{%
  \institution{University of California - Berkeley}
  \city{Berkeley}
  \state{CA}
  \country{USA}}
\email{h.pasandi@berkeley.edu}

\author{Faith Parastar}
\orcid{}
\affiliation{%
  \institution{Independent Researcher}
  \city{}
  \state{}
  \country{}}


\renewcommand{\shortauthors}{Hannah B. Pasandi}

\begin{abstract}
  
  As we transition into the era of B5G/6G networks, the promise of seamless, high-speed connectivity brings unprecedented opportunities and challenges. Among the most critical concerns is the preservation of location privacy, given the enhanced precision and pervasive connectivity of these advanced networks. This paper systematically reviews the state of knowledge on location privacy in B5G/6G networks, highlighting the architectural advancements and infrastructural complexities that contribute to increased privacy risks. The urgency of studying these technologies is underscored by the rapid adoption of B5G/6G and the growing sophistication of location tracking methods. We evaluate current and emerging privacy-preserving mechanisms, exploring the implications of sophisticated tracking methods and the challenges posed by the complex network infrastructures. Our findings reveal the effectiveness of various mitigation strategies and emphasize the important role of physical layer security. Additionally, we propose innovative approaches, including decentralized authentication systems and the potential of satellite communications, to enhance location privacy. By addressing these challenges, this paper provides a comprehensive perspective on preserving user privacy in the rapidly evolving landscape of modern communication networks.
\end{abstract}


\maketitle
\vspace{-0.15 in}
\section{Introduction}

\begin{figure}
\centering
	\includegraphics[width=0.49\textwidth, height=1.8 in]{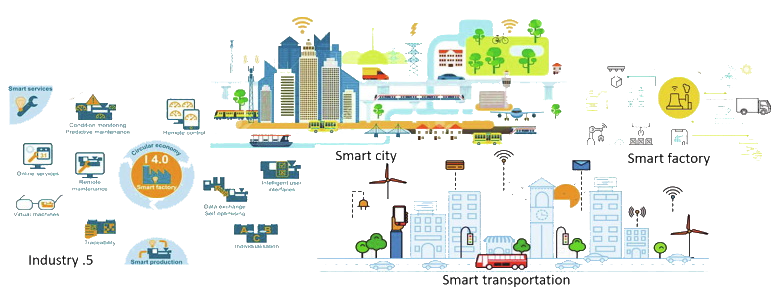}
	\caption{B5G/6G Services and Applications}
	\label{fig:services}
\vspace{-0.15in}
\end{figure}

We are transitioning into a connected and intelligent era with the commercialization of 5G and the ongoing standardization of 6G. These next-generation cellular networks represent not just incremental updates, but a profound technological revolution. They promise seamless connectivity with significantly higher data rates, expanded coverage, and support for a wide array of innovative applications such as Augmented/Virtual Reality (AR/VR), smart health, smart transportation, the Internet of Things (IoT), and the Industrial Internet of Things (IIoT)~\cite{ferrag2018security,rafique2023securemed}. Furthermore, 6G anticipates incorporating satellite communication integration, Artificial Intelligence (AI) at the network edge, Green network with more efficiency of energy consumption, and data transmission at terabytes per second by using a heterogeneous access network. Figure.~\ref{fig:services}illustrates a comprehensive list of services and applications. This transformative potential stems from various groundbreaking technologies deployed in these networks~\cite{dao2021survey,sandeepa2022survey}.

The importance of studying B5G/6G lies in the need to understand and address the complex privacy concerns associated with the enhanced precision and ubiquitous connectivity of B5G/6G. The urgency is driven by the rapid deployment of these networks and the growing capabilities of adversaries to exploit location data, which can lead to severe privacy breaches and security threats. This study provides a comprehensive analysis of the current state of location privacy, identifies emerging threats, and evaluates the effectiveness of existing and novel privacy-preserving mechanisms. By doing so, it lays the groundwork for future research and development efforts aimed at ensuring user privacy in the context of advanced communication networks. The insights gained from this study will help inform the design of robust privacy frameworks, guide policymakers, and enhance the security and trustworthiness of next-generation networks. Given the critical nature of location data and the potential for its misuse, it is imperative to focus on this aspect to protect user privacy and maintain trust in emerging technologies.

In 5G, key technologies such as Network Function Virtualization (NFV) and Software Defined Networking (SDN) play important roles in reshaping network architecture and operations. NFV enables the virtualization of network functions, allowing them to run on standard hardware, thus enhancing scalability and flexibility. SDN, on the other hand, decouples the control plane from the data plane, enabling centralized network management and dynamic resource allocation. Multiple Input Multiple Output (MIMO) technology in 5G significantly enhances spectral efficiency and network capacity by utilizing multiple antennas for transmission and reception, thereby enabling higher data rates and improved coverage~\cite{schmitt2021pretty, xie2022ng, porambage2021roadmap, nayak20216g, rahman2022deep}. Looking ahead to B5G and 6G, further advancements are anticipated, leveraging technologies such as terahertz frequencies, intelligent surfaces, and quantum communication. Terahertz frequencies offer the potential for ultra-high data rates and extremely low latency, enabling a new generation of applications. Intelligent surfaces, also known as reconfigurable intelligent surfaces (RIS), manipulate electromagnetic waves to enhance wireless communication, improve coverage, and mitigate interference. Quantum communication, leveraging the principles of quantum mechanics, promises unparalleled security and ultra-fast data transmission rates, revolutionizing the landscape of secure communication in B5G/6G networks~\cite{iavich2021novel, norrman2016protecting}. 

Despite the technological advancements and the AI~\cite{1,2,3,4,5,6} role in 5G and beyond era~\cite{yang2022federated}, concerns regarding privacy and security persist, particularly in the context of location data. For example, the integration of FL and AI also expands the vulnerability surface. The decentralized nature of FL introduces potential attack vectors such as model poisoning, where malicious entities could corrupt the learning process. Similarly, AI systems, despite their robustness, can be susceptible to adversarial attacks that exploit weaknesses in their training data or algorithms. As B5G/6G networks become more reliant on these technologies, it is crucial to develop advanced security measures to mitigate these risks and ensure the integrity and reliability of the network~\cite{lin2024split}.

Therefore, the initial privacy issues in B5G/6G networks arise from the fundamental architectural differences (Section~\ref{sec:architecture}) introduced in their design. Unlike their predecessors, these advanced networks incorporate decentralized architectures, heterogeneous network components, and advanced technologies like Massive MIMO, mmWave communications, and edge computing. \textit{These infrastructural complexities significantly increase the precision and volume of location data collected, thus amplifying the risk of privacy breaches. Consequently, the increased granularity and accuracy of location tracking in B5G/6G networks expose user data to sophisticated tracking methods and unauthorized access, highlighting the urgent need for robust privacy-preserving mechanisms to protect sensitive information} (Section~\ref{sec:location}).


To provide a better understanding of the current and emerging 5/6G technologies with respect to privacy concerns and mitigation, this paper delves into the intricate landscape of user privacy with the focus on location privacy in B5G/6G networks, exploring the challenges and proposed solutions in protecting user privacy in the evolving technological landscape. Through a systematization of knowledge, we shed light on the complexities of this critical issue and provide insights for future research and development efforts~\cite{sandeepa2022survey}.


\textbf{Contributions.} This paper makes the following key contributions to the study of architectural and infrastructural complexities, and location privacy in B5G/6G networks:

\begin{itemize}
    \item   \textbf{Comprehensive Analysis of Location Privacy Challenges.} We provide an in-depth analysis of the unique location privacy challenges posed by B5G/6G networks. This includes the infrastructural complexities, increased data precision, and the evolving threat landscape that significantly impact user privacy.

   \item\textbf{Evaluation of Privacy-Preserving Mechanisms} We systematically review and evaluate existing and emerging privacy-preserving mechanisms. This evaluation covers techniques such as anonymization, pseudonymization, and homomorphic encryption, assessing their effectiveness in the context of B5G/6G.

    \item\textbf{Architectural and Infrastructural Complexities:} We explore the architectural and infrastructural complexities introduced by B5G/6G networks, such as the transition from centralized to distributed architectures, the integration of heterogeneous network components, and the deployment of advanced technologies like Massive MIMO, mmWave communications, and edge computing. These complexities are analyzed in terms of their implications for network performance, scalability, and location privacy.

   \item \textbf{Innovative Solutions for Enhanced Privacy:} We propose novel approaches to enhance location privacy, including decentralized authentication systems and the integration of satellite communications. These solutions aim to address the identified privacy challenges and provide robust protection against unauthorized tracking and data breaches.

    \item \textbf{Guidance for Future Research and Policy Development:} By identifying the gaps and limitations in current privacy-preserving strategies, this paper offers valuable insights for future research directions and policy development. We emphasize the need for regulatory oversight and user awareness to create a comprehensive framework for protecting location privacy in next-generation networks.
\end{itemize}


\begin{figure}[!t]
\centering
	\includegraphics[width=0.45\textwidth, height=2 in]{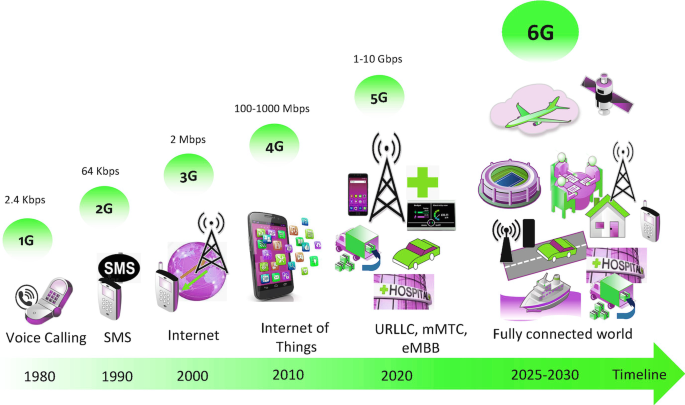}
	\caption{Cellular Network Evolution}
	\label{fig:5gArch}
\vspace{-0.15in}
\end{figure}
\section{Mobile Network Evolution}
The transition from 4G to 5G and beyond (Figure.~\ref{fig:5gArch}) marks a significant leap in mobile network technology, bringing substantial enhancements in connectivity and speed alongside new challenges in location privacy. While 4G networks enabled applications requiring moderate location accuracy, such as navigation and social media, 5G networks offer much more precise location tracking. This improved precision, achieved through higher frequency bands like millimeter-wave (mmWave) and advanced positioning techniques, is essential for applications such as AR, smart cities, and autonomous vehicles. However, this increased accuracy also raises significant privacy concerns, as more detailed location data becomes available and potentially exploitable.

The technological advancements in 5G, including Software Defined Networking (SDN), Network Function Virtualization (NFV), and edge computing, contribute to the precision of location services while introducing new privacy challenges. SDN and NFV enable flexible and efficient management of network resources, crucial for maintaining high accuracy in dynamic environments. Edge computing processes data closer to the user, reducing latency and enhancing real-time location data processing. This local processing capability is critical for applications requiring instant location updates, such as emergency services and drone navigation. However, the increased granularity of location data processed at the edge also amplifies the risks of unauthorized access and privacy breaches, necessitating robust privacy-preserving mechanisms.

B5G/6G networks are set to enhance location capabilities further with advanced technologies, while simultaneously posing greater privacy risks. B5G aims to refine the precision and reliability of 5G, while 6G will integrate artificial intelligence (AI) and machine learning (ML) to provide unprecedented levels of location accuracy and context-awareness. Innovations such as terahertz (THz) frequencies, intelligent surfaces, and quantum communication are expected to significantly improve location tracking. These advancements will enable ultra-high-definition location services for applications like remote surgery and immersive virtual environments. However, the ability to track user locations with such high precision increases the potential for privacy infringements, making it essential to develop and implement advanced privacy-preserving technologies.

As mobile networks evolve from 4G to B5G and 6G, the complexity of managing and securing location data intensifies, highlighting the urgent need for robust privacy measures. The deployment of new technologies requires extensive infrastructure upgrades and introduces new challenges in data privacy and security. Higher frequency bands, such as mmWave and THz, necessitate a denser network of small cells, complicating the network architecture and increasing the potential for location data breaches. The enhanced capability to track user locations with high precision exacerbates privacy concerns, necessitating advanced security protocols and privacy-preserving techniques. The integration of AI and ML in 6G networks, while improving location accuracy, adds layers of complexity in ensuring that location data is handled securely and ethically, emphasizing the need for comprehensive privacy frameworks.

\begin{figure*}[!t]
\centering
	\includegraphics[width=0.8\textwidth, height=3.18 in]{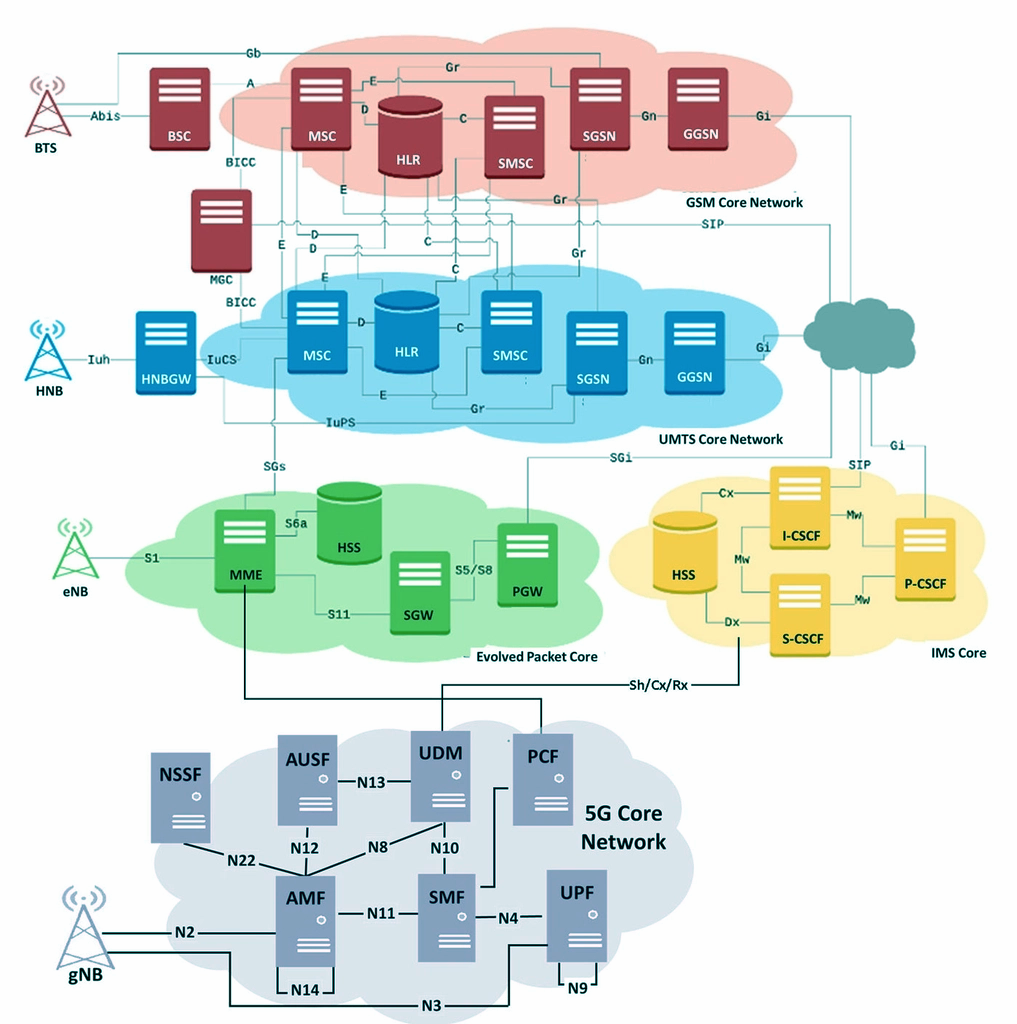}
	\caption{Architectural Evolution of Cellular Networks}
	\label{fig:arch-ev}
\vspace{-0.15in}
\end{figure*}

\vspace{-0.15 in}

\section{Architectural Differences and Implications}
\label{sec:architecture}
The architectural differences between successive generations of cellular networks, such as 4G, 5G, and upcoming B5G/6G, have profound implications for network performance, scalability, and security. These differences stem from advancements in technology, changes in network architecture, and evolving user demands (Figure.~\ref{fig:arch-ev}). One notable architectural shift is the transition from centralized to distributed architectures. In 4G networks, core network functions are typically centralized, leading to potential bottlenecks and scalability challenges. In 4G networks, localization and positioning methods primarily rely on a combination of cellular network infrastructure, satellite systems, and Wi-Fi access points. With methods such as cell-ID based positioning, WiFi positioning system (WPS), Assisted-GPS (A-GPS), 4G network could locate users with accuracy of 10m and 1km~\cite{del2017survey,checa2020location,becker2022could}.

In contrast, 5G networks embrace a more distributed architecture, leveraging concepts like NFV and SDN to decentralize network functions. the NGC decouples the user plane (or data plane) from the control plane (CP). This function, which is known as CUPS2 (Control \& User Plane Separation), was first introduced in 3GPP release 14. An important characteristic of this function being that, in case of a traffic peak, you can dynamically scale the CP functions without affecting the user plane operations, allowing deployment of UP functions (UPF) closer to the RAN and UE to support use cases like Ultra Reliable low latency Communication (URLLC). This decentralization improves scalability, flexibility, and reliability, enabling more efficient resource allocation and dynamic network management. Another significant architectural difference lies in the increased deployment of small cells and the utilization of mmWave frequencies in 5G and beyond. Small cells, including enhanced network capacity and coverage, particularly in dense urban environments. mmWave frequencies enable higher data rates and lower latency but require denser deployment due to shorter propagation distances. This architectural shift necessitates new infrastructure planning and deployment strategies to optimize coverage and capacity while mitigating interference and propagation challenges. However, the localization accuracy with mmWave transmissions can be on the order of 10 m or less outdoors and 3 m indoors, in part thanks to cooperative gNBs and massive-MIMO with angle-of-arrival (AoA) and angle-of-departure measurements~\cite{del2018whitepaper}. Figure~\ref{fig:arch-ev} illustrates the architectural evolution of cellular networks, from 2G to 5G.

Edge computing represents a paradigm shift in network architecture, bringing computation and data storage closer to the network edge, but at the same time causing more concise location data. By processing data locally at the edge of the network, edge computing reduces latency, enhances real-time responsiveness, and improves scalability. This architectural evolution is particularly relevant for latency-sensitive applications such as AR/VR and autonomous vehicles, where real-time data processing is critical. 

\begin{figure}[t]
\centering
	\includegraphics[width=0.4\textwidth, height=2.7 in]{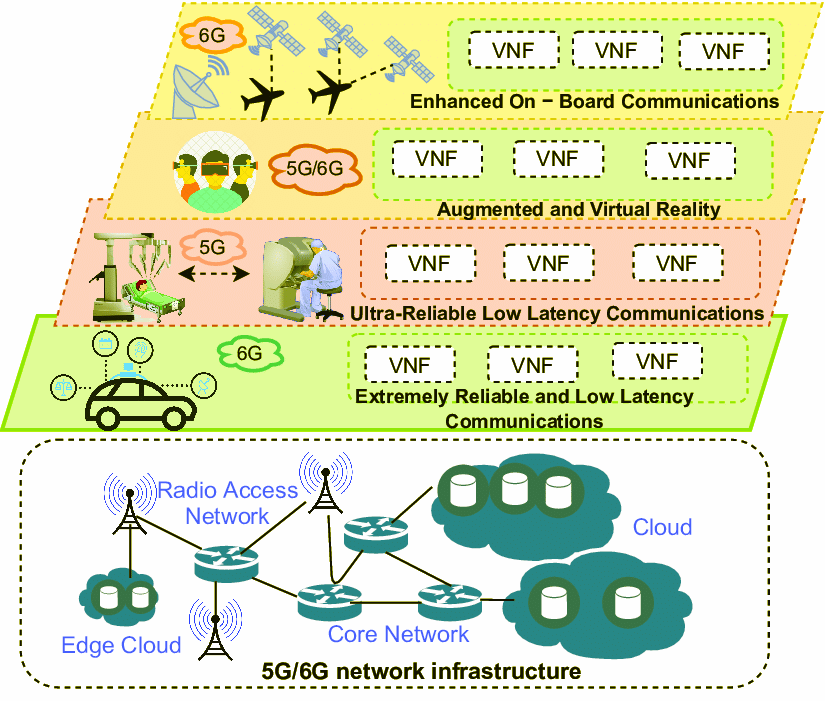}
	\caption{B5G/6G Network Slicing}
	\label{fig:slicing}
\vspace{-0.15in}
\end{figure}

Network slicing is a key architectural concept in 5G and beyond, enabling the creation of virtualized network instances tailored to specific use cases or service requirements. By dynamically allocating network resources based on application demands, network slicing enables efficient resource utilization, improved QoS, and enhanced network customization. However, implementing network slicing requires robust orchestration and management frameworks to ensure isolation, security, and performance guarantees for each slice. Moreover, the proliferation of network slices adds complexity to network operations and monitoring, necessitating sophisticated management and automation solutions. Figure.~\ref{fig:slicing} shows different levels of network slicing.

the deployment of small cells and the integration of edge computing introduce new opportunities for location-based services and applications. However, they also raise concerns about the collection, processing, and sharing of location data. Small cells, being deployed in closer proximity to users, enable more precise localization and tracking, which can compromise user privacy if not properly regulated. Additionally, edge computing platforms may aggregate and analyze location data from multiple sources, creating potential privacy risks if sensitive information is exposed or misused~\cite{tomasin2021location}.

The 6G network core is anticipated to build upon the foundation established by 5G but with significant enhancements and new features to meet the evolving demands of future communication technologies, such as integrated AI/ML and Sustainability and Energy Efficiency considerations. with seamless integration of terrestrial networks with non-terrestrial networks such as satellites, HAPs, and UAVs enhances the accuracy and availability of location data. This integration ensures continuous and precise location tracking even in remote areas where terrestrial infrastructure is sparse. The fusion of multiple sources of location data from different types of networks improves overall localization accuracy and reliability. Further Enhanced satellite communication provides global positioning capabilities with higher precision, enabling accurate location tracking anywhere on the planet. This is particularly useful for applications requiring precise geolocation, such as navigation in remote areas, environmental monitoring, and global logistics~\cite{ferrag2018security,rafique2023securemed,sandeepa2022survey}.

\vspace{-0.15 in}

\section{Location Privacy \& It's challenges in B5G/6G Networks}
\label{sec:location}
Location data privacy is paramount in today's digital age due to its pervasive nature and potential for misuse. With the advent of advanced mobile technologies like 5G and the imminent arrival of B5G/6G networks, the granularity and accuracy of location tracking have reached unprecedented levels. Every time an individual accesses the internet, makes a call, or uses location-based services, their precise whereabouts are logged, creating a digital footprint that can reveal sensitive information about their daily routines, habits, and even personal preferences. This wealth of location data is highly coveted by various stakeholders, including advertisers, marketers, and even malicious actors seeking to exploit it for financial gain or nefarious purposes. Misuse of location data poses severe risks, such as unauthorized tracking and profiling, which can lead to stalking, identity theft, and physical harm. These dangers underscore the critical need for robust privacy protection measures to protect individuals' sensitive information and prevent malicious exploitation~\cite{jiang2021location}.

Despite the proliferation of privacy-preserving techniques and regulatory frameworks aimed at protecting user data, the privacy of location remains a persistent challenge. One key reason for this ongoing struggle is the inherent tension between privacy and utility in the digital ecosystem. While users demand seamless and personalized services that leverage their location data, they also expect robust privacy protections to prevent unauthorized access or misuse. Balancing these competing interests is no easy feat, as evidenced by the numerous instances of data breaches, privacy violations, and regulatory lapses that continue to plague the industry. Moreover, the rapid pace of technological innovation and the ever-expanding array of interconnected devices further complicate efforts to maintain privacy in an increasingly interconnected world~\cite{haddad2023blockchain,babaghayou2023safety}.

Furthermore, the evolving threat landscape and the emergence of novel privacy risks underscore the need for continuous improvement in location data privacy measures. Despite the existence of encryption protocols, anonymization techniques, and data minimization practices, vulnerabilities persist, leaving users vulnerable to exploitation and manipulation. Moreover, the proliferation of IoT devices, smart sensors, and wearables only exacerbates these challenges by introducing new vectors for data collection and analysis. As such, addressing the privacy implications of location data requires a multifaceted approach that combines technical innovation, regulatory oversight, and user education to create a more resilient and privacy-centric digital ecosystem~\cite{sandeepa2022survey,khan2020survey,wang2024gees}.

Maintaining user privacy is challenging in cellular networks, both past and present as it is not a primary goal of the architecture. In order to authenticate users for access and billing purposes, networks use globally unique client identifiers. Like- wise, the cellular infrastructure itself must always “know” the location of a user in order to minimize latency when providing connectivity~\cite{schmitt2021pretty}. New advancements of B5G/6G provide us with seamless- high speed connection, new services and applications with unimaginable quality; But, they also have a dark side. These benefits come with the price of possible privacy breaches. In following, we mentioned some outstanding and beneficial features of B5G/6G that affect our privacy more than before.

\subsection{Increased Precision and Volume}
With features such as mmWave, The integration of terrestrial and non-terrestrial networks, including satellites and high-altitude platforms, B5G/6G enhances global connectivity but also expands the scope of location tracking. B5G/6G networks enable ultra-precise location tracking with accuracies down to the decimeter level. While this is beneficial for applications like autonomous vehicles and AR, it also raises concerns about intrusive surveillance and unauthorized tracking. This high precision of location data increases the risk of misuse by malicious actors, advertisers, or unauthorized third parties. Detailed location information can be exploited for targeted advertising, stalking, or profiling without the user's consent. Also, Cross-border data flows between terrestrial and non-terrestrial networks raise jurisdictional and legal challenges for protecting location privacy. Different regulatory frameworks and data protection laws across countries may result in gaps or inconsistencies in privacy protections~\cite{anant2023introduction,shaham2020privacy,lin2024single}.
\vspace{-0.15in}
\subsection{Advance Tracking Methods}
As we mentioned in previous sections, cellular networks use GNS data to locate the users. However with the advancements in B5G/6G this data are more accurate and detailed. A-GPS enhances traditional GPS by using additional assistance data from cellular networks to improve satellite acquisition and positioning accuracy. It enables faster location fixes and works well in urban environments or areas with obstructed GPS signals. WPS leverages WiFi signals from nearby access points to determine the location of a device. By triangulating signals from multiple WiFi access points, WPS can provide accurate positioning, especially in indoor environments where GPS signals may be weak or unavailable. Also, AI-driven analytics can analyze historical location data to predict user behavior and preferences. While this enables personalized services and recommendations, it also raises concerns about predictive surveillance and intrusion into personal lives. As during the corona virus pandemic lots of new tracking methods proposed based on ML methods, while they were aiming health of society, they also could endanger privacy of people~\cite{shubina2020survey,ciaparrone2020deep,bagayogo2024new}. 
\vspace{-0.15in}
\subsection{Navigating the Complexity of B5G/6G Infrastructure}
As we advance towards B5G and 6G networks, the infrastructural complexity of cellular networks significantly increases, introducing new challenges and opportunities for maintaining location privacy. The evolution from centralized to decentralized architectures, the integration of heterogeneous network components, and the incorporation of advanced technologies like Massive MIMO, mmWave communications, and edge computing contribute to this complexity. These enhancements aim to provide higher data rates, lower latency, and improved connectivity, but they also introduce additional layers of vulnerability. Ensuring robust location privacy in such a multifaceted and dynamic environment requires a thorough understanding of these infrastructural complexities and the development of sophisticated security measures to mitigate associated risks. Some of main such complexities are: 

\subsubsection{ Multi Access Access Network.}
\begin{figure}[t]
\centering
	\includegraphics[width=0.4\textwidth, height=1.6 in]{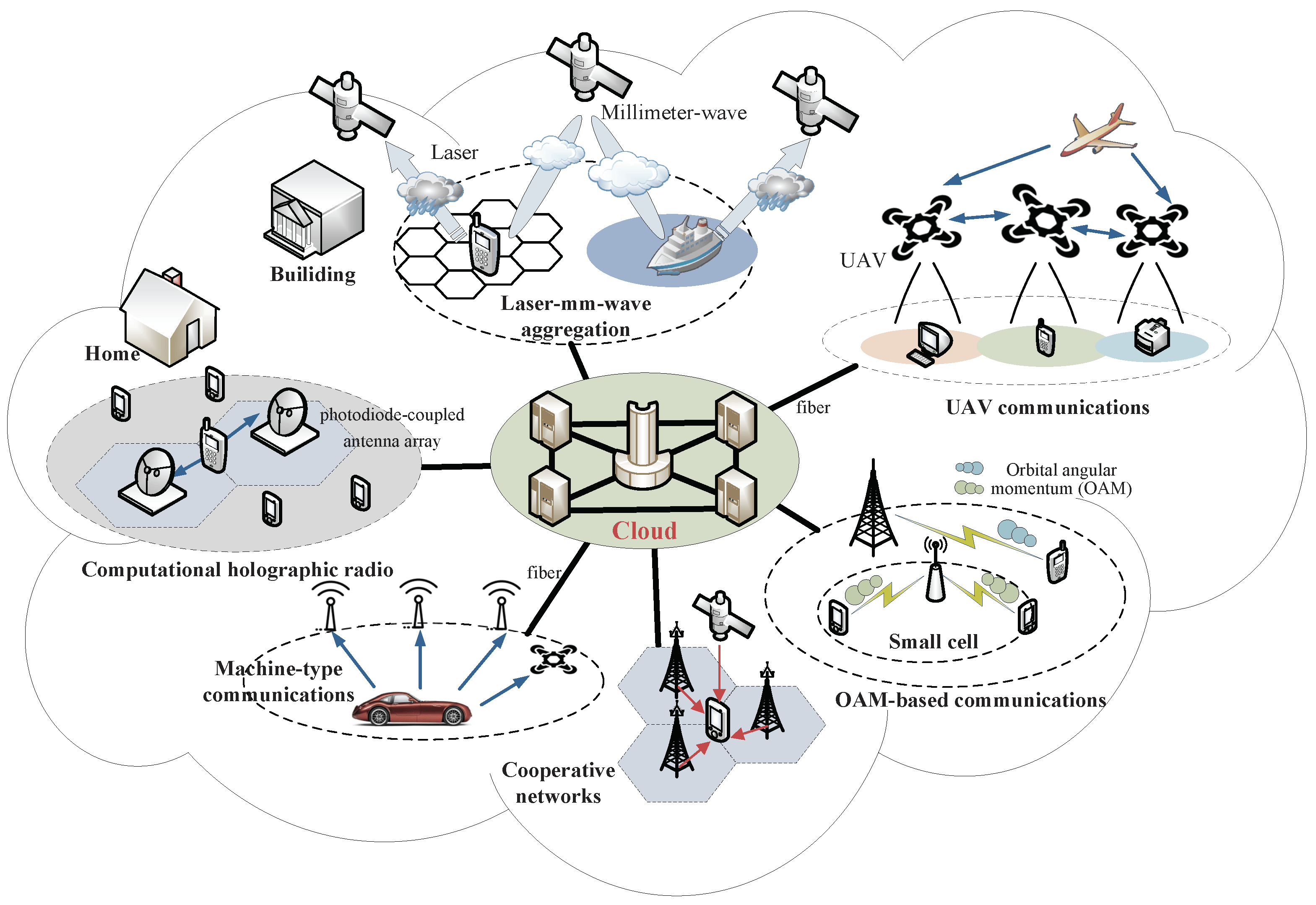}
	\caption{B5G/6G Access Network}
	\label{fig:access}
\vspace{-0.15in}
\end{figure}

Modern networks encompass a variety of access technologies such as cellular, WiFi, Bluetooth, and satellite communication. Each of these technologies has unique infrastructure and protocols for gathering location data, resulting in fragmentation and challenges with interoperability. The access network in B5G/6G is highly complex, incorporating terrestrial and non-terrestrial elements such as drones, satellites, and connected vehicles to ensure seamless, ubiquitous connectivity. This multi-layered infrastructure poses significant challenges in terms of integration, management, and security, necessitating advanced technologies and robust protocols to maintain efficient and secure communication. Integrating location data from these diverse access networks while maintaining consistency and accuracy is a significant challenge, particularly when it comes to preserving privacy~\cite{kazmi2023security,sandeepa2022survey,tomasin2021location,schmitt2021pretty,lin2024single,wang2024gees}. Figure~\ref{fig:access} shows the complexity of access network in B5G/6G.

\vspace{-0.2 in}
\subsubsection{Edge and Fog Computing.}
Mobile edge computing (MEC) is essential to the low-latency use cases that form an integral part of B5G/6G. For example, organisers of a sports event may want to provide slow motion videos or augmented reality content to people in the stadium. This could be achieved by storing the content related to the game in MEC servers closest to the stadium. Similarly, a factory may collect data from sensors, vehicles, wearable devices and drones used in the factory and process these locally helping it to operate more efficiently~\cite{hu2015mobile,mourtzis2022design,liyanage2021driving}.
While the servers used in mobile edge computing may reside closer to network elements to reduce latency, they are essentially providing a cloud or hosting service. MEC not only decreases latency but also lowers the energy required for data transmission, which is one of important aspects of B5G/6G. Additionally, it can optimize resource allocation and computational tasks dynamically, leading to more efficient use of network and device power. 
The ability to process data in the network edge in which it has been generated, could obviate the need to send the data further into or outside the network, reducing wider circulation of personal data than is necessary.
It may be important to distinguish between the content data which is stored locally by the MEC provider to provide the service and the communications data which would remain subject to the usual licence conditions and legal requirements~\cite{zhang2020edge,ranaweera2021survey}.

By processing data closer to the user inherently precise location information involved in optimizing performance and reducing latency. This proximity increases the risk of location data being intercepted or misused, as the data is handled at multiple edge nodes rather than centralized servers. Additionally, the distributed nature of MEC creates more potential points of vulnerability, making it harder to secure all nodes effectively. This expansion of the attack surface heightens the potential for unauthorized access and exploitation of sensitive location information~\cite{wang2024gees}.

\subsubsection{Network Slicing and Virtualization.}
Network slicing enables the creation of multiple virtual networks on a shared physical infrastructure, with each network customized for specific applications or service needs. This architectural flexibility allows for dynamic resource allocation and personalized network services, but it also complicates the management and isolation of location data. To ensure privacy in virtualized network environments, robust isolation mechanisms and strong authentication protocols are necessary. To offer multiple heterogeneous services from a single platform which is flexible in nature, the flexibility is provided by a network slicing mechanism in which one physical network is divided into multiple logical networks to offer these services simultaneously. 

These slices use generic resources to serve users and specific applications. SDN and NFV play a significant role in slicing to make the 5G network more scalable and reliable. Network slicing deals with three prominent use cases: the first enhances mobile broadband (eMBB) for those applications which require high bandwidth to offer Ultra-HD video communication. This use case increases the traffic load on mobile networks. The second massive machine type communication (mMTC) offers connectivity between millions of devices. This use case does not increase traffic but increases the magnitude of mobile networks. The last service is URLLC which applies to connectivity, remote surgery, industry 4.0, etc. in vehicles to everything (V2X). This use case needs very low latency connectivity~\cite{dangi2022ml}.
The network slicing architectural framework is based on three categories:
\textit{RAN slicing}, 
\textit{Core Network slicing} and \textit{End-to-End (E2E) slicing}.
Each network slice creation is done in four phases: softwarization, virtualization, orchestration, and management.

\vspace{-0.15in}
\subsubsection{IoT and M2M communication.}
The proliferation of IoT devices and machine-to-machine (M2M) communications generates vast amounts of location data from sensors, actuators, and connected devices. These devices often operate autonomously and may transmit location information without explicit user consent. As these devices become interconnected, they can communicate with nearby devices and stream data continuously. This will create a highly sophisticated digital environment, making it virtually possible to track every user action. Consequently, B5G/6G networks will pave the way for a fully automated smart environment that could encompass a vast population. Managing privacy in IoT deployments requires implementing privacy-by-design principles, data minimization strategies, and secure communication protocols~\cite{sandeepa2022survey}. For the secure collaboration between nodes and the detection of potential security threats within the network, it is beneficial to establish trust rankings for nodes. This is especially useful in defense and healthcare networks, where the security of the network is heavily reliant on the trustworthiness of its nodes~\cite{shafi2023intelligent}.

\subsubsection{Roaming.}
One of the approaches that can help to the resilience of cellular networks is roaming which facilitates mobile network operators to use the infrastructure of each other when needed, e.g., in case its own infrastructure is not functional or does not suffice to serve with the required service levels.
Preserving location privacy and keep the communication of UEs safe during roaming is essential. Designing a secure and efficient authentication protocol for roaming users in the mobile network is a challenging. In order to secure communication over an insecure channel, authentication schemes have been proposed. The main weakness of the existing authentication protocols is that attackers have the ability to impersonate a legal user at any time. In addition, the existing protocols are vulnerable to various kind of cryptographic attacks such as insider attack, bit flipping attack, forgery attacks, DoS attack, unfair key agreement and cannot provide user’s anonymity~\cite{madhusudhan2020mobile,kang2021lightweight,mafakheri2021smart,weedage2023resilience} and in case of a successful attack the attacker can achieve the serial ID of user and track it even after the attack is over. 
\subsubsection{Cloud-based Services.}
Many location-based services rely on cloud-based infrastructure and data centers for storage, processing, and analysis of location data. While cloud computing offers scalability and accessibility, it also raises concerns about data sovereignty, data ownership, and third-party access to sensitive location information. Securing location data in cloud environments requires encryption, data anonymization, and strict access controls to prevent unauthorized access or data breaches.

\subsection{A Common Scenario: False Base Station Detection Techniques}
The term "false base stations" (FBS) refers to attacks where an imposter base station pretends to be a legitimate one. This issue is currently being studied by the SA3 working group, as documented in TR 33.809 [1]. In 5G, an FBS typically acts as a "man-in-the-middle" (MitM) or a very covert jammer. A significant vulnerability exposed by FBSs is that the initial phases of network entry, which occur before the implementation of 5G security protocols, are especially susceptible to many of the attacks. For instance, attacks that involve replaying altered versions of broadcast channels can severely impact all the terminals in a cell, either disrupting their network connection or forcing them to operate in a compromised state. Therefore, it is crucial to develop methods that enable user equipment (UE) to verify the legitimacy of a base station before exchanging unauthenticated messages by incorporating a base station authentication factor. To achieve this, Physical Layer Security (PLS) could be utilized for localization by a UE as a form of soft authentication~\cite{chorti2022context,anant2023introduction,kazmi2023security,irram2022physical}.

\vspace{-0.15in}
\section{Privacy-Preserving Mechanisms}
\begin{figure*}
\centering
	\includegraphics[width=0.6\textwidth, height=2 in]{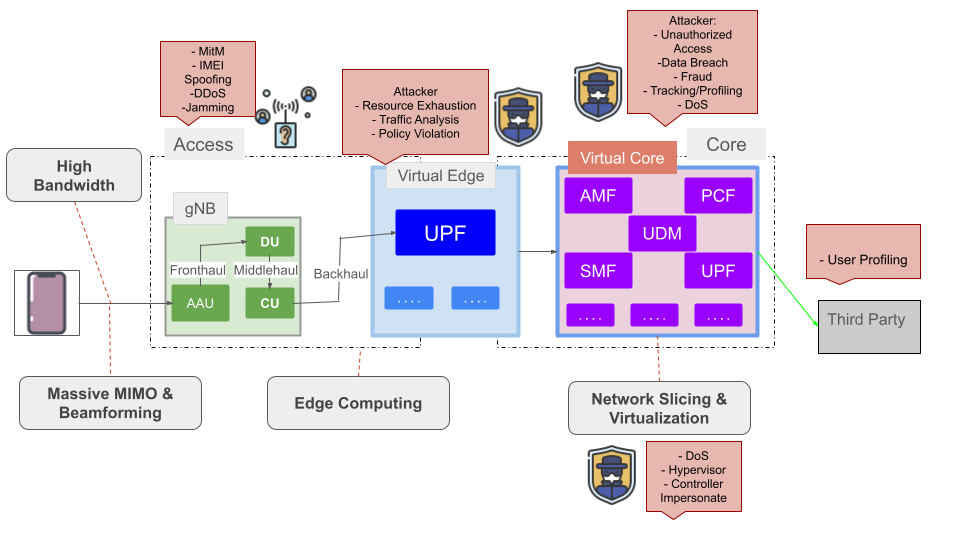}
	\caption{Attacks Toxonomy in B5G/6G}
	\label{fig:taxonomy}
\vspace{-0.15in}
\end{figure*}

As mobile networks evolve into B5G/6G, the volume and precision of location data being generated increases dramatically, making the protection of user privacy more critical than ever. The advancement in technologies, such as mmWave and MIMO in 5G, and the integration of AI and satellite communications in 6G, have amplified the need for robust privacy-preserving mechanisms. This section delves into the various strategies and technologies currently employed to preserve location privacy, evaluates their effectiveness, and discusses emerging solutions aimed at addressing the unique challenges posed by the latest generation of mobile networks. Figure.\ref{fig:taxonomy} illustrates the taxonomy of various attacks targeting location privacy within B5G/6G networks. It categorizes the different types of attacks that exploit vulnerabilities at various layers of the network infrastructure

\subsection{Current Advanced Privacy Techniques Solutions}
A variety of technological, governmental, and legal approaches are now being used to protect the integrity and confidentiality of location data in order to offer location privacy. 

\subsubsection{Anonymization and Psuedonymization.}
Homomorphic encryption, differential privacy, and pseudonymization are key methods used to anonymize network data, including location data, during transmission. Homomorphic encryption enables computations on encrypted data without needing decryption, thereby maintaining data privacy even during processing. Differential privacy adds random noise to data sets, ensuring that individual data points cannot be re-identified while allowing for aggregate data analysis. Pseudonymization replaces identifiable information with pseudonyms, reducing the risk of re-identification and enhancing privacy. These methods are often used in combination to provide robust protection against privacy breaches. For instance, differential privacy can be applied to pseudonymized data to further obscure individual identities, while homomorphic encryption can secure data during both storage and computation phases. Together, these techniques offer a layered approach to data protection, protecting sensitive information from unauthorized access and ensuring compliance with stringent data privacy regulations~\cite{anant2023introduction,haque2021anonymous,sicari20205g,lin2024single}. This combination of advanced cryptographic and anonymization techniques is crucial for maintaining the privacy and security of location data in modern communication networks.

\subsubsection{Authentication.}
Authentication is essential to a comprehensive strategy for protecting location privacy. By verifying that only authorized users and devices have access to location data, employing sophisticated authentication techniques, and incorporating secure protocols, organizations can greatly reduce the risks of unauthorized tracking and data breaches. Alongside other privacy-enhancing technologies and regulatory measures, strong authentication systems are vital for ensuring user privacy in current and future communication networks~\cite{haque2021anonymous,sicari20205g,dawar2024enhancing,ma2023uav}.

\subsubsection{Location Masking.}
Techniques for location masking typically involve obfuscating precise location coordinates or substituting them with less accurate or more generic identifiers to enhance user privacy. These strategies aim to balance the utility of location data with robust privacy protections, ensuring individuals cannot be easily tracked or profiled. One common method is differential privacy, which adds random noise to location data, hiding true locations within a range of possible values. Another approach is spatial cloaking, where precise coordinates are replaced with broader areas like grid cells or regions, maintaining data relevance for regional analyses while obscuring individual details. Temporal cloaking reduces the frequency of location updates, preventing detailed tracking of movement patterns. Advanced cryptographic techniques, such as homomorphic encryption, allow for processing location data while it remains encrypted, ensuring sensitive information is not exposed. Dynamic location cloaking adjusts the level of obfuscation based on context, offering higher protection for sensitive areas. These masking strategies maintain the usefulness of location data for aggregate analysis and operational purposes while significantly reducing the risk of unwanted tracking and profiling, adhering to privacy regulations and user expectations~\cite{sung2020zipphone}.

\subsubsection{Privacy-by-Design.}
Privacy-by-design principles are promoted when location-based services are designed and developed with privacy considerations integrated from the beginning. This proactive approach ensures that privacy is not an afterthought but a fundamental aspect of the development process. By incorporating privacy-enhancing features such as data minimization, developers can limit the amount of personal data collected and processed to only what is necessary for the intended purpose, thereby reducing the potential for data breaches and misuse. Purpose limitation ensures that data is used exclusively for the reasons explicitly stated to users, preventing unauthorized secondary uses that could compromise privacy. Additionally, user-centric controls empower individuals to manage their privacy settings actively, giving them greater control over their personal information. These controls can include options for users to opt-in or opt-out of data collection, specify data retention periods, and control who can access their data. By embedding these principles into the core architecture of location-based services, developers can significantly mitigate privacy risks, enhance user trust, and ensure compliance with stringent regulatory frameworks such as GDPR and CCPA. This holistic approach not only protects user privacy but also fosters a culture of transparency and accountability in the development and deployment of location-based technologies~\cite{masouros2023towards}.

\subsubsection{Improving Awareness.}
People can make informed choices regarding their privacy preferences when they are told about the dangers and consequences of location tracking and receive guidance on privacy-enhancing techniques. User-friendly privacy controls, privacy literacy programs, and awareness efforts increase public knowledge of location privacy issues while promoting responsible data stewardship.

\subsubsection{Regulation.}
Regulation refers to a set of rules, policies, and laws established by government authorities or regulatory bodies to govern and control various aspects of 5G technology. These regulations are designed to ensure that the 5G operates in a safe, fair, and orderly manner, with the goal of protecting the interests of consumers, fostering competition, and addressing potential risks or challenges~\cite{bauer2018roles,morgado2018survey}. Regulations can cover a wide range of areas, including: Spectrum allocation, network security, privacy, competition, and environmental impact.
Legal requirements for protecting location data and preserving user privacy are established by regulatory frameworks such as the Federal Communications Commission (FCC) and California Consumer Privacy Act (CCPA) in the US and the General Data Protection Regulation (GDPR) in Europe. To ensure compliance with these requirements, privacy measures must be put in location, privacy impact assessments must be carried out, and data processing procedures must be transparent and accountable~\cite{anant2023introduction,bauer2022regulation, morgado2018survey}.

\subsubsection{ML-based Solution.}
ML algorithms can analyze vast amounts of data to identify patterns and anomalies that may indicate unauthorized access or malicious activities. For instance, ML can improve the effectiveness of anomaly detection systems by continuously learning from new data and evolving threats, making it harder for attackers to evade detection. Additionally, ML can be used in adaptive and contextual authentication systems, where it can dynamically adjust security measures based on the user's behavior and context, significantly enhancing security while maintaining user convenience~\cite{pandey2024privacy,ferrag2018security,rafique2023securemed}.

\begin{figure*}
\centering
	\includegraphics[width=0.6\textwidth, height=2 in]{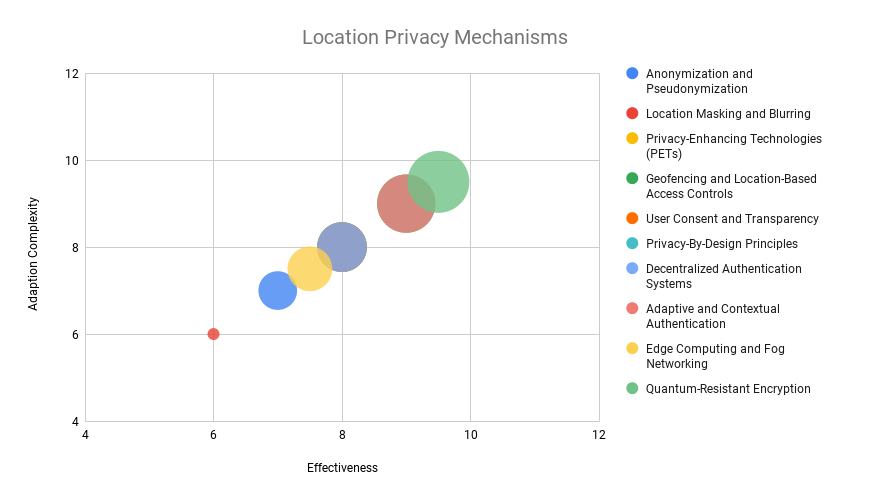}
	\caption{Location Privacy Mechanisms}
	\label{fig:mechanisms}
\vspace{-0.2 in}
\end{figure*}

\vspace{-0.2 in}
\subsection{Emerging Solutions for B5G/6G}
Preserving privacy in B5G and 6G networks is crucial as these networks are expected to handle an unprecedented amount of sensitive data from a vast array of connected devices such as IoX and VANET networks. To preserve privacy to high amount of detailed data, new solutions have been proposed by the community. Some of these methods, such as homomotphic encryption are the same as previous versions of network. Encryption methods are vital to provide data security in hierarchy of network and have a long history. But new techniques to cope with high bitrate and energy effeciency standards of 6G are necessary. Figure.~\ref{fig:mechanisms} illustrates a comparison between the effectiveness and complexity of adoption of various privacy-enhancing solutions in B5G/6G networks. The figure presents a visual assessment of different solutions like differential privacy, blockchain, multi-party computation, and secure multi-access edge computing, highlighting their relative strengths in ensuring user privacy versus the complexity involved in implementing these technologies. The axes of the graph show the balance between the solution's effectiveness in protecting privacy and the complexity or effort required to adopt and integrate these solutions into existing and future network infrastructures.

\subsubsection{Adaptive and Contextual Authentication.}  
Adaptive and contextual authentication systems enhance security by dynamically adjusting authentication requirements based on various factors, such as user behavior, device characteristics, and environmental context. These systems continuously analyze real-time data to assess the risk level associated with each authentication attempt. For instance, if a user attempts to log in from an unusual location or at an atypical time, the system might prompt for additional verification steps, such as multi-factor authentication, to ensure the legitimacy of the access request. This approach balances security and usability by tightening security controls in high-risk situations while minimizing friction in low-risk scenarios. By leveraging machine learning and advanced analytics, adaptive and contextual authentication provides a robust defense against unauthorized access, making it particularly suitable for applications requiring high security, such as financial services and sensitive data management~\cite{bumiller2023understanding,zhao2021open,le2022novel}.

\subsubsection{Federated Learning.}
 FL is a decentralized machine learning approach that is particularly well-suited for addressing the data privacy and computational challenges of 6G networks. In the context of 6G, FL allows multiple devices, such as smartphones and IoT devices, to collaboratively train a machine learning model without sharing their raw data.  Instead of sending raw data to a central server, only the model updates are shared, significantly reducing the risk of data breaches and preserving user privacy. FL is particularly beneficial for applications like predictive maintenance in smart cities or personalized healthcare, where sensitive data is involved.  FL can also be integrated with edge computing, enabling local data processing and reducing latency, which is crucial for real-time applications like autonomous driving, augmented reality, and smart healthcare. By optimizing network resources through the prediction of traffic patterns and management of network load, FL can enhance the QoS and reduce operational costs. Moreover, FL enables the development of personalized services by training models on-device using user-specific data, which enhances user experience in applications like personalized recommendations, health monitoring, and smart home automation~\cite{yang2022federated,priya20246g}.

FL also contributes to robust security mechanisms, such as anomaly detection and fraud prevention, by collaboratively detecting potential security threats across the network without compromising individual user data. Its application in federated authentication enhances security protocols, making systems more resilient to attacks. Additionally, FL improves the performance and intelligence of IoT applications in smart cities, industrial automation, and environmental monitoring. Despite these benefits, FL faces challenges such as communication overhead, heterogeneous device capabilities, data and model security, and ensuring model convergence across diverse datasets and devices. Addressing these challenges is essential to fully realize the potential of FL in 6G networks.

\subsubsection{Homomorphic Encryption.} 
 Homomorphic encryption approaches are useful in scenarios where data needs to be processed by third-party services without compromising privacy, such as in cloud computing and secure data sharing among organizations. But as Quantom computing becomes reality of B5G/6G, quantum-resistant encryption methods are being developed to protect data against quantum attacks. These encryption methods are still in the research phase and may require significant computational resources~\cite{narayanan2008robust}.

\subsubsection{Differential Privacy.} 
Differential privacy introduces noise into datasets to obscure individual entries, making it difficult to identify specific individuals while still allowing for meaningful aggregate data analysis. This technique is valuable for data analytics and machine learning applications where it is crucial to derive insights from data without exposing personal information. For instance, differential privacy can be used in smart grids to analyze energy consumption patterns without revealing individual household usage~\cite{wu2020location}.

\subsubsection{Blockchain and Distributed Ledger Technologies.} 
These technologies provide a decentralized and secure way to manage data transactions and access. By using blockchain, data can be securely shared and accessed in a transparent manner, with tamper-evident records ensuring data integrity. This is particularly useful for applications like supply chain management, where multiple parties need to access and verify transaction records without compromising privacy~\cite{soni2023blockchain,rafique2023securemed,priya20246g}.

\subsubsection{Multi-Party Computation (MPC).}
MPC allows multiple parties to collaboratively compute a function over their inputs while keeping those inputs private. This technique ensures that sensitive data is not revealed to other parties during the computation process. MPC can be applied in scenarios such as collaborative research, where institutions need to share data insights without exposing proprietary information~\cite{luz2024secure}.

\subsubsection{Secure Multi-Access Edge Computing (MEC).} 
MEC brings computation closer to the data source, reducing latency and enhancing privacy by processing data locally on edge devices rather than sending it to central servers. This approach minimizes the exposure of sensitive data to potential breaches during transmission and central processing. It is particularly beneficial for applications like autonomous vehicles and industrial IoT, where real-time processing and privacy are critical~\cite{liyanage2021driving}.

These new solutions collectively enhance the ability of B5G and 6G networks to protect user privacy while enabling advanced data processing and analytics, ensuring that the benefits of these next-generation networks can be fully realized without compromising security and trust. Figure~\ref{fig:mechanisms} illustrating a comparison between these solution's effectiveness and complexity of adaption. 

Anonymization and pseudonymization methods are generally effective at preventing direct identification of individuals but can be vulnerable to re-identification attacks when combined with auxiliary data, offering an effectiveness rated at 70\%. Location masking and blurring techniques reduce the precision of location data to protect privacy but degrade the quality of location-based services, with an effectiveness rated at 60\%. Privacy-enhancing technologies (PETs) are highly effective at protecting privacy by minimizing data exposure, though they can be computationally intensive, achieving an effectiveness rating of 90\%~\cite{narayanan2008robust}.

Geofencing and location-based access controls are effective at controlling access based on location but rely on accurate location data and can be spoofed, with an effectiveness rated at 80\%. User consent and transparency mechanisms ensure that users are aware of and consent to data usage, though they can suffer from consent fatigue, also rated at 80\% effectiveness. Privacy-by-design principles are highly effective when implemented from the ground up but require significant organizational change, rated at 90\%. Decentralized authentication systems improve security and privacy through decentralization, though they can be complex and challenging to scale, rated at 80\% effectiveness. Adaptive and contextual authentication methods are very effective at ensuring security based on context, though resource-intensive, rated at 90\%. Edge computing and fog networking enhance privacy by processing data closer to the source but introduce new security challenges due to their distributed nature, rated at 75\%. Quantum-resistant encryption promises very high security against future quantum attacks, though still largely experimental, rated at 95\% effectiveness~\cite{alshantti2024privacy,bashir2024progress,narayanan2008robust}. 
\vspace{-0.2in}
\section{Discussion}

\subsection{Physical Layer Security for B5G/6G Privacy}

Physical layer security is critical in preserving location privacy within B5G/6G networks. The physical layer is where communication signals are transmitted and received, making it a prime target for various attacks. Key attacks that exploit vulnerabilities at this layer include:

\textbf{Man-in-the-Middle (MitM) Attacks:} In these attacks, an adversary intercepts and potentially alters the communication between the user and the network without either party's knowledge. By doing so, attackers can gain access to sensitive location data and other personal information. MitM attacks can be particularly damaging in scenarios where location data is used for sensitive applications such as financial transactions or health services.

\textbf{Jamming:} This involves deliberate interference with communication signals, causing degradation in service quality. Attackers can use jamming to disrupt legitimate connections, forcing devices to reconnect to malicious base stations or fallback to less secure protocols. This can lead to the exposure of location data and other personal information.

\textbf{Fingerprinting:} Attackers analyze the unique characteristics of signals emitted by devices to identify and track them over time. By creating a profile of a device's signal patterns, attackers can track a user's movements and activities, compromising their privacy. For example, if an intruder successfully fingerprints a 5G base station, they can gather a variety of information that can be used for various purposes, some potentially malicious. This information includes the Cell Identifier (Cell ID), Public Land Mobile Network (PLMN) ID, Tracking Area Code (TAC), Global Cell ID (GCI), and Physical Cell ID (PCI), which provide detailed insights into the network's layout and geographic scope. They can also collect data on the frequency bands and specific channels used, supported technologies like 4G LTE and 5G NR, and base station capabilities such as MIMO and carrier aggregation. Additionally, they can map neighboring cells and assess signal strength and quality metrics. 
The potential for misuse of this information includes detailed mapping of the network for planning attacks, creating targeted jamming attacks, setting up rogue base stations for MITM attacks, and exploiting weaknesses in handover procedures. It can also facilitate location tracking of users and manipulation of network behavior~\cite{nazerian2021passive,li2019passive}. 

\textbf{Fake Base Stations (FBS):} Also known as rogue base stations, these are unauthorized devices that mimic legitimate base stations to deceive UE into connecting to them. Once connected, these fake base stations can capture location data and other sensitive information. FBS attacks highlight a significant vulnerability, especially during the initial network entry phase before the establishment of secure communication protocols.
To mitigate these threats, it is crucial to implement robust physical layer security measures. Techniques such as advanced encryption, secure authentication, signal obfuscation, and the use of spread spectrum technologies can enhance security at the physical layer. Ensuring that devices and base stations authenticate each other before exchanging sensitive information can also prevent unauthorized access.
\vspace{-0.2 in}
\subsection{Decentralized Systems for Privacy Protection}
Distributed systems and technologies like blockchain offer significant potential for enhancing location privacy in B5G/6G networks. These technologies help decentralize authentication processes and data storage, reducing the risks associated with centralized systems. For example, distributed authentication approach involves authenticating users and devices across multiple, independent nodes rather than relying on a single centralized authority. This reduces the risk of a single point of failure and makes it more difficult for attackers to compromise the authentication process. By using distributed authentication, the system can ensure that even if one node is compromised, the overall integrity and security of the network remain intact~\cite{soni2023blockchain,nyangaresi2023privacy}.

Further, blockchain provides a decentralized and tamper-proof ledger for recording authentication transactions. It ensures transparency and security by recording each authentication attempt in an immutable ledger. This can help verify user identities without revealing sensitive location data, thereby enhancing privacy. Blockchain can also support smart contracts to enforce privacy policies automatically and securely. Storing user identity separately from location data means that even if an attacker gains unauthorized access to one set of data, they cannot easily correlate it with other personal information. This separation minimizes the potential damage from data breaches and enhances overall privacy. For instance, identity verification can be handled by one part of the network, while location data processing occurs elsewhere, reducing the risk of comprehensive data leaks~\cite{priya20246g,luz2024secure}. 
\vspace{-0.2 in}
\subsection{Robust Security with Satellite Communication}

Satellite communication introduces a robust layer of security that can enhance location privacy. Satellites are physically more difficult to access and tamper with compared to terrestrial infrastructure. This inherent security makes it significantly harder for attackers to conduct MitM attacks, deploy fake base stations, or perform jamming.
Satellite communications often employ sophisticated and proprietary encryption methods that add a significant layer of security. These encryption techniques make it exceedingly difficult for attackers to intercept and decipher location data. Additionally, satellite networks can dynamically change encryption keys and use advanced cryptographic protocols to further enhance security.
Satellites provide consistent and secure communication across vast geographic areas, ensuring that users in remote or underserved regions can benefit from enhanced privacy protections. This global coverage also enables more reliable and resilient communication, reducing the chances of location data being exposed through weak terrestrial connections~\cite{beyaz2024satellite}.

\vspace{-0.15 in}
\subsection{User-Centric Privacy Awareness \& Regulatory Oversight}

Improving user awareness and strengthening regulatory oversight are crucial components in preserving location privacy. Users need to be informed about the risks associated with location data and the measures they can take to protect their privacy. Educating users on the importance of regularly reviewing and updating their privacy settings on devices and applications. Providing clear instructions and tools to manage these settings can empower users to take control of their privacy.
Offering guidelines on safe internet practices, such as recognizing phishing attempts, avoiding suspicious networks, and using encrypted connections. Promoting the use of privacy-enhancing tools, such as VPNs and secure communication apps, can also help protect location data.

Regulatory entities must also recognize the advancements in cellular networks as critical infrastructure and update privacy regulations accordingly. Ensuring that privacy regulations keep pace with technological advancements in B5G/6G networks. 
Implementing strict enforcement mechanisms to hold MNOs accountable for breaches of privacy by conducting regular audits, imposing penalties for non-compliance, and requiring transparency in data handling practices.
Also, encouraging the adoption of privacy-by-design principles in the development and deployment of new technologies can be done by regulations. This approach ensures that privacy considerations are integrated into the design and operation of systems from the outset, rather than being added as an afterthought. By combining technological, educational, and regulatory efforts, we can create a comprehensive approach to preserving location privacy in the evolving landscape of B5G/6G networks~\cite{benlloch2023distributed,nokia}.
\vspace{-0.2 in}
\section{Conclusion \& Overlook}
The evolution from 4G to B5G/6G networks marks a significant technological leap, bringing about revolutionary changes in connectivity, speed, and application potential. However, these advancements also introduce new challenges, particularly regarding the privacy of location data. As these networks provide higher precision in tracking and a vast increase in connected devices, the potential for privacy breaches escalates. In this paper, we explored the critical importance of location privacy in the context of B5G/6G networks. We discussed the architectural advancements and their implications, highlighting how new technologies such as NFV, SDN, MIMO, and others, while enhancing network capabilities, also pose privacy risks. We identified and evaluated various location privacy mechanisms, recognizing the limitations and effectiveness of current solutions and provide about adapting complexity of these methods. We shouldn't overlook the crucial role of ML in both presenting challenges and offering solutions within the context of B5G/6G location privacy. ML algorithms can be exploited for advanced tracking and analysis, potentially compromising user privacy by predicting and inferring location patterns with high accuracy. However, the same technology can also be harnessed to enhance privacy protections. ML can enable sophisticated anomaly detection systems that identify and mitigate suspicious activities, implement adaptive and contextual authentication mechanisms, and optimize data anonymization techniques. Thus, while ML introduces new vulnerabilities, it also offers powerful tools to fortify privacy measures, necessitating a balanced and strategic approach in its deployment within next-generation networks.

We also emphasized the need for robust physical layer security and advanced tracking methods, noting the potential threats from attacks like fingerprinting, man-in-the-middle, jamming, and fake base stations. The discussion extended to the benefits of distributed authentication systems and the potential role of satellite communication in improving location privacy. Finally, we underscored the importance of regulatory measures and increasing user awareness to mitigate privacy risks. As we move towards a highly connected and automated world, preserving location privacy remains an open and pressing challenge. Continued research and innovative solutions are imperative to ensure that the benefits of B5G/6G networks are realized without compromising user privacy.
\vspace{-0.15 in}
\section*{Acknowledgments}
The authors would like to acknowledge the use of AI-based tools, specifically ChatGPT~\cite{chatgpt} developed by OpenAI, for assistance in correcting grammatical errors and improving the writing of this paper. The AI tool was utilized to revise the text throughout the paper, particularly focusing on correcting any typos, grammatical errors, and awkward phrasing.

\vspace{-0.15 in}

\bibliographystyle{plain}
\small{
    \bibliography{sample-base.bib}
}

\end{document}